\documentstyle[12pt,axodraw]{article}

\catcode`@=12
\begin{document}
\thispagestyle{empty}
\begin{flushright} 
UCRHEP-T315\\ 
September 2001\
\end{flushright}
\vspace{0.5in}
\begin{center}
{\LARGE \bf Double Threefold Degeneracies\\ for Active and Sterile Neutrinos\\}
\vspace{1.5in}
{\bf Ernest Ma$^a$ and G. Rajasekaran$^b$\\}
\vspace{0.2in}
{\sl $^a$ Physics Department, University of California, Riverside, 
California 92521\\}
\vspace{0.1in}
{\sl $^b$ Institute of Mathematical Sciences, Chennai (Madras) 600113, India\\}
\vspace{1.5in}
\end{center}
\begin{abstract}\
We explore the possibility that the 3 active (doublet) neutrinos have 
nearly degenerate masses which are split only by the usual seesaw mechanism 
from 3 sterile (singlet) neutrinos in the presence of a softly broken $A_4$ 
symmetry. We take the unconventional view that the sterile neutrinos may be 
light, i.e. less than 1 keV, and discuss some very interesting and novel 
phenomenology, including a connection between the LSND neutrino data and 
solar neutrino oscillations.
\end{abstract}
\newpage
\baselineskip 24pt

Present experimental data \cite{atm,solar,lsnd} indicate that neutrinos 
oscillate.  Hence they should have small nonzero masses and mix with one 
another.  On the other hand, since neutrino oscillations only measure 
the difference of mass squares, the possibility that all 3 active neutrinos 
are nearly degenerate in mass should not be overlooked \cite{amr}.  There 
are two canonical ways of making $m_\nu$ nonzero.  One is through the small 
vacuum expectation value (VEV) of a Higgs triplet \cite{svgr,msmrs}.  The 
other is through the addition of 3 heavy singlet neutral fermions (usually 
considered as right-handed neutrinos $N_R$).  In that case, a Dirac mass 
$m_D$ linking the left-handed doublet neutrinos $\nu_L$ with $N_R$ as well 
as a Majorana mass $M$ for $N_R$ are allowed.  Combining the two mechanisms, 
the following mass matrix
\begin{equation}
{\cal M}_{\nu N} = \left( \begin{array} {c@{\quad}c} m_0 & m_D \\ m_D & M 
\end{array} \right)
\end{equation}
is obtained.  The eigenvalues are simply $m_0 - m_D^2/M$ and $M$.  Without 
$m_0$ (which comes from the VEV of the Higgs triplet), this is just the 
famous seesaw mechanism \cite{seesaw} for a small neutrino mass.  The singlet 
$N_R$ is too heavy to be detected experimentally, unless \cite{ma01} 
$m_D$ comes from a different Higgs doublet with a suppressed VEV, in which 
case $M$ may in fact be only a few TeV or less and become observable at 
future colliders.  Using the model of Ref.\cite{ma01}, it has also been shown 
\cite{mr} that the possibly large observed discrepancy of the muon anomalous 
magnetic moment \cite{g-2} may be explained, provided that the 3 active 
neutrinos are in fact nearly degenerate in mass, in order not to conflict 
with the present experimental bound on $\tau \to \mu \gamma$.

In this paper we consider the case where both $m_D$ and $M$ are small, but 
$m_D$ is still less than $M$ by perhaps an order of magnitude.  This is in 
contrast to the pseudo-Dirac scenario \cite{koli}, i.e. $m_0, M << m_D$, in 
which case neutrino oscillations would be maximal between active and sterile 
species, in disfavor with the most recent data \cite{atm,solar}.  We also 
supplement our model with a discrete $A_4$ symmetry \cite{maraj,acl} which 
maintains the separate degeneracies of the 3 active and 3 sterile neutrinos. 
This $A_4$ is then broken spontaneously and softly to allow for realistic 
charged-lepton masses as well as neutrino mass differences as in 
Ref.\cite{maraj}.  The new idea here is that the 3 sterile neutrinos could be 
light and help to account for the LSND data \cite{lsnd} as shown below. 

Before discussing the theoretical reasons for $m_0$, $m_D$, and $M$ to be 
small, consider first the phenomenology of such a possibility.  The 3 active 
neutrinos $\nu_e$, $\nu_\mu$, $\nu_\tau$ are now each a linear combination of 
6 light neutrino mass eigenstates.  With $m_D$ less than $M$ by an order of 
magnitude, the mixing of $N$ with $\nu$ is still small; hence the presumably 
large mixings among the 3 active neutrinos themselves are sufficient to 
explain the atmospheric \cite{atm} and solar \cite{solar} neutrino data.  
This leaves the LSND data \cite{lsnd} to be explained by the mixing of $\nu$ 
with $N$.

Consider Eq.~(1) as a $6 \times 6$ matrix with $m_0$ and $m_D$ representing 
$3 \times 3$ unit matrices, as required by the $A_4$ symmetry.  The soft 
breaking of $A_4$ means that $M$ may differ slightly from the unit matrix, 
so that in the basis under which it is diagonal, ${\cal M}_{\nu N}$ is 
given by
\begin{equation}
{\cal M}_{\nu N} = \left[ \begin{array} {c@{\quad}c@{\quad}c@{\quad}c@{\quad}c
@{\quad}c} m_0 & 0 & 0 & m_D & 0 & 0 \\ 0 & m_0 & 0 & 0 & m_D & 0 \\ 
0 & 0 & m_0 & 0 & 0 & m_D \\ m_D & 0 & 0 & M_1 & 0 & 0 \\ 
0 & m_D & 0 & 0 & M_2 & 0 \\ 0 & 0 & m_D & 0 & 0 & M_3 \end{array} \right].
\end{equation}
The $\nu_e, \nu_\mu, \nu_\tau$ basis is now rotated into the $\nu'_{1,2,3}$ 
basis, and we may assume whatever pattern is suitable for explaining the 
atmospheric and solar neutrino data.  To be specific, consider bimaximal 
mixing, i.e.
\begin{equation}
\left[ \begin{array} {c} \nu'_1 \\ \nu'_2 \\ \nu'_3 \end{array} \right] = 
\left[ \begin{array} {c@{\quad}c@{\quad}c} 1/\sqrt 2 & 1/2 & -1/2 \\ 
-1/\sqrt 2 & 1/2 & -1/2 \\ 0 & 1/\sqrt 2 & 1/\sqrt 2 \end{array} \right] 
\left[ \begin{array} {c} \nu_e \\ \nu_\mu \\ \nu_\tau \end{array} \right].
\end{equation}
Then the eigenstates of ${\cal M}_{\nu N}$ are
\begin{equation}
\nu_i = \nu'_i \cos \theta_i - N_i \sin \theta_i, ~~~ 
S_i = \nu'_i \sin \theta_i + N_i \cos \theta_i,
\end{equation}
where $\sin \theta_i \simeq m_D/M_i$, corresponding to the eigenvalues 
$m_0 - m_D^2/M_i$ and $M_i$ respectively.  Since $M_1 \simeq M_2 \simeq M_3$ 
is still assumed, we have
\begin{equation}
\Delta m_{ij}^2 = \left( m_0 - {m_D^2 \over M_i} \right)^2 - \left( m_0 - 
{m_D^2 \over M_j} \right)^2 \simeq {m_\nu m_D^2 \over M^3} \Delta M_{ij}^2,
\end{equation}
where $M = (M_1 + M_2 + M_3)/3$ and $m_\nu = m_0 - m_D^2/M$.

Consider now the effect of $S_i$ on $\nu_\mu \to \nu_e$ oscillations.  The 
well-known expression for this probability is given by
\begin{equation}
P(\nu_\mu \to \nu_e) = - 4 \sum_i U_{\mu i} U_{e i} \sum_{j>i} U_{\mu j} 
U_{e j} \sin^2 \left( {\Delta m_{ij}^2 L \over 4 E} \right).
\end{equation}
For the $L/E$ values appropriate to the LSND experiment, $\Delta m_{ij}^2$ 
is effectively zero between $\nu_i$ and $\nu_j$.  Naively, we might expect 
the contribution from $\Delta m_{ij}^2$ between $S_i$ and $\nu_j$, i.e. 
$M^2 - m_\nu^2$ to be dominant, but that turns out to be negligible. 
Specifically,
\begin{equation}
\sum_{j=4,5,6} U_{\mu j} U_{e j} = - \sum_{i=1,2,3} U_{\mu i} U_{e i} = 
{1 \over 2 \sqrt 2}(\sin^2 \theta_1 - \sin^2 \theta_2) \simeq {1 \over 2 
\sqrt 2} \left( {\Delta m_{21}^2 \over m_\nu M} \right) \simeq 0,
\end{equation}
where Eq.~(5) has been used.  Hence the main contribution to Eq.~(6) is 
actually coming from $i=4$ and $j=5$, i.e.
\begin{equation}
P(\nu_\mu \to \nu_e) = {1 \over 2} \sin^2 \theta_1 \sin^2 \theta_2 \sin^2 
\left( {\Delta M_{12}^2 L \over 4 E} \right).
\end{equation}
This means that the neutrino mass difference being probed by the LSND 
experiment is that between $S_1$ and $S_2$, and \underline {not} that 
between $\nu_e$ or $\nu_\mu$ and $S_i$.  To understand this interesting 
new phenomenon, we note that if the $M_i$'s were equal, then $\nu_e$, 
$\nu_\mu$, $\nu_\tau$ would all be exactly degenerate in mass and there 
could not be any $\nu_\mu \to \nu_e$ oscillation; thus any such effect 
must be proportional to the difference in the $M_i$'s and not to the 
difference between $M$ and $m_\nu$.

To fit the LSND data, we take $\Delta M_{21}^2 \simeq 1$ eV$^2$ and 
$\sin^2 \theta_1 \simeq \sin^2 \theta_2 \simeq 0.06$, so that 
$\sin^2 2 \theta_{eff} \simeq 1.8 \times 10^{-3}$.  
In that case, Eq.~(5) relates them to $\Delta m_{21}^2$ which may be as 
large as $1.2 \times 10^{-4}$ eV$^2$ as indicated by the solar data.  
We then obtain $m_\nu/M \simeq 2 \times 10^{-3}$.  If we take $m_\nu \simeq 
0.2$ eV, then $M \simeq 0.1$ keV.  Taking $\Delta m_{32}^2$ to be as small as 
$1.2 \times 10^{-3}$ eV$^2$ as indicated by the atmospheric data, we 
obtain $\Delta M_{32}^2 \simeq 10$ eV$^2$ and $\sin^2 2 \theta_{eff} \simeq 
3.6 \times 10^{-3}$ for $\nu_\mu \to \nu_\tau$ oscillations in the CHORUS 
\cite{chorus} and NOMAD \cite{nomad} experiments, which are just beyond 
their exclusion boundaries.

We now give the theoretical details of our model.  In addition to $A_4$, we 
define 2 global U(1) symmetries:  $U(1)_L$ and $U(1)_S$.  Under these 
symmetries, the leptons transform as follows: \cite{maraj} $(\nu_i,l_i)_L 
\sim (\underline {3}, 1, 0)$, $l_{1R} \sim (\underline {1}, 1, 0)$, $l_{2R} 
\sim (\underline {1}', 1, 0)$, $l_{3R} \sim (\underline {1}'', 1, 0)$, 
$N_{iR} \sim (\underline {3}, 1, 1)$.  There are 4 scalar doublets: $\Phi_i 
= (\phi^+_i, \phi^0_i) \sim (\underline {3}, 0, 0)$, $\eta = (\eta^+,\eta^0) 
\sim (\underline {1}, 0, 1)$, and 1 scalar triplet: $\xi = (\xi^{++}, \xi^+, 
\xi^0) \sim (\underline {1}, -2, 0)$.  The soft terms $\Phi_i^\dagger \eta$ 
break $U(1)_S$ and $A_4$, $\xi^\dagger \Phi_i \Phi_j$ break $U(1)_L$ and 
$A_4$, $N_i N_j$ break $U(1)_L$, $U(1)_S$, and $A_4$, hence they may all be 
naturally small \cite{thooft}.  Note that the smallness of $M$ is protected 
by \underline {both} $U(1)_L$ and $U(1)_S$.  Assuming $m_\xi^2$ and 
$m_\eta^2$ to be positive and large, we then obtain $\langle \xi^0 \rangle$ 
and $\langle \eta^0 \rangle$ to be small \cite{msmrs,ma01} for the terms 
$m_0$ and $m_D$ respectively.

Since the 3 sterile neutrinos have masses of order 0.1 keV and their decay 
lifetimes (through their mixing with the active neutrinos) are much greater 
than the age of the Universe, they would overclose the Universe unless their 
relic abundance is greatly reduced.  This may be achieved in our scenario 
because $N_{iR}$ have no gauge interactions and the only Yukawa couplings 
they have are given by
\begin{equation}
{\cal L}_{int} = f \bar N_{iR} (\nu_i \eta^0 - l_i \eta^+) + h.c.
\end{equation}
Hence they decouple from the standard-model particles at the scale $M_\eta$ 
which we take to be 1 TeV, and whereas $\nu_{e,\mu,\tau}$ are heated 
by the subsequent annihilations of nonrelativistic particles, $N_{iR}$ are 
not \cite{oss}.  Thus the number densities of the latter are greatly 
suppressed until the onset of electroweak symmetry breaking.  However, the 
mixing of $N_{iR}$ with $\nu_i$ is then large enough to produce the former 
through active-sterile neutrino oscillations and we are faced again with the 
abundance problem, together with the nucleosynthesis bound of the effective 
number of neutrinos $n_\nu$ which is restricted to be less than 4 \cite {bbn}. 
To evade these cosmological problems, we need $N_{iR}$ to decay quickly as 
proposed previously \cite{decay}, but at the expense of the unnatural fine 
tuning of parameters.

In conclusion, we have constructed in this paper a specific model of 3 
active (doublet) and 3 sterile (singlet) neutrinos, which are separately 
threefold degenerate in mass approximately.  We find the new and interesting 
result that neutrino oscillations between active species are governed by 
the mass differences among the 3 lighter neutrinos and the parallel mass 
differences [see Eq.~(5)] among the 3 heavier neutrinos, but not the 
mass difference between the two groups.  If bimaximal mixing is assumed for 
explaining the atmospheric and solar neutrino oscillations (using the 
matter-enhanced solution for the latter), then a possible explanation of 
the LSND data is the nonnegligible mixing ($\sin^2 \theta \simeq 0.06$) 
between active and sterile neutrinos.  This would then imply 
sterile neutrino masses of order 0.1 keV (if active neutrino masses are 
around 0.2 eV) and $\Delta m^2 \simeq 10$ eV$^2$ for the CHORUS and NOMAD 
experiments with $\sin^2 \theta_{eff} \simeq 3.6 \times 10^{-3}$, which is 
just barely consistent with their exclusion limits.  New data from future 
long-baseline and medium-baseline neutrino-oscillation experiments will be 
decisive in confronting this possible 3+3 scenario.

\vskip 0.3in
This work was supported in part by the U.~S.~Department of Energy
under Grant No.~DE-FG03-94ER40837.  G.R. also thanks the UCR Physics 
Department for hospitality and acknowledges INSA (new Delhi) for support.

\newpage
\bibliographystyle{unsrt}

\end{document}